\documentclass[12pt]{iopart}
\usepackage{graphicx}
\usepackage{color}

\begin{document}

\title{Charmonium Transverse Momentum Distribution in High Energy Nuclear Collisions}

\author{Zebo Tang$^1$, Nu Xu$^{2,3}$, Kai Zhou$^4$ and Pengfei Zhuang$^4$}

\address{$^1 $ Department of Modern Physics, University of Science and Technology of China, Hefei 230026, China\\
         $^2 $ Key Laboratory of Quark and Lepton Physics (MOE) and Institute of Particle Physics, Central China Normal University, Wuhan 430079, China\\
         $^3 $ Nuclear Science Division, Lawrence Berkeley National Laboratory, Berkeley, CA 94720, USA\\
         $^4 $ Physics Department, Tsinghua University and Collaborative Innovation Center of Quantum Matter, Beijing 100084, China}

\date{\today}

\begin{abstract}
The Charmonium transverse momentum distribution is more sensitive to the nature of the hot QCD matter created in high energy nuclear collisions, in comparison with the yield. Taking a detailed transport approach for charmonium motion together with a hydrodynamic description for the medium evolution, the cancelation between the two hot nuclear matter effects, the dissociation and the regeneration, controls the charmonium transverse momentum distribution. Especially, the second moment of the distribution can be used to differentiate between the hot mediums produced at SPS, RHIC and LHC energies.

\end{abstract}

\section{Introduction}
\label{s1}

It is widely accepted that, there exists a phase transition from hadron matter to quark matter when the temperature or baryon density of a Quantum
Chromodynamics (QCD) system is high enough~\cite{shuryak}. This deconfinement phase transition has been simulated by lattice QCD at finite temperature~\cite{latqgp1,latqgp2}. Experimentally, the only way to realize a high temperature QCD system in laboratories is through high energy nuclear collisions. From the Au+Au collisions with colliding energy $\sqrt {s_{NN}}$ = 200 GeV at the Relativistic Heavy Ion Collider (RHIC)~\cite{rhicexp1,rhicexp2,rhicexp3,rhicexp4,rhicthe} and the Pb+Pb collisions with $\sqrt {s_{NN}}$ = 2.76 TeV at the Large Hadron Collider (LHC)~\cite{lhcexp1,lhcexp2,lhcexp3}, the critical temperature for the phase transition from hadron gas to quark-gluon plasma (QGP) looks like to be reached in the collisions. The most important problem in the study of QGP in heavy ion collisions is how to signal its formation in the early stage. Due to the rapid expansion of the colliding system, the temperature of the fireball drops down with time, and the QGP, if it is created in the initial stage, is only an intermediate state and can
not be directly detected in the final state. Therefore, we need sensitive signatures to access the properties
of the hot medium. The $J/\psi$ suppression proposed by Matsui and Satz~\cite{satz86} has long been considered as an essential probe for the deconfinement phase transition.

Since quarkonia are tightly bound states of a pair of heavy quarks $Q$ and $\bar Q$, their dissociation temperatures $T_d$, calculated by non-relativistic~\cite{pot1,pot2} and relativistic~\cite{guo} potential models or lattice QCD simulations, are above the critical temperature $T_c$ of the deconfinement of light quarks. Therefore, the produced quarkonia are statically suppressed only in hot medium with temperature $T>T_d$ where the color screening radius becomes shorter than the quarkonium size. However, the hot nuclear matter effects on quarkonia include not only the color-screening induced suppression, but also the quarkonium regeneration~\cite{thews,munzinger,rapp}. The number of heavy quarks created in the initial stage of heavy ion
collisions increases substantially with colliding energy. There are more than 10 $c\bar c$ pairs produced in a
central Au+Au collision at RHIC, and the
number is probably over 100 in heavy ion collisions at LHC~\cite{gavai}. In this case, the recombination of those uncorrelated heavy quarks $Q$ and $\bar Q$ becomes the second source of the quarkonium production. Obviously, this regeneration will enhance the quarkonium yield. The regeneration approach for
$J/\psi$ in heavy ion collisions has been widely discussed with different models, such as the thermal creation on the hadronizaton hypersurface
according to statistic law~\cite{munzinger}, the coalescence mechanism~\cite{gorenstein,gerco}, and the kinetic model which
considers both initial production and continuous regeneration of quarkonium yields~\cite{thews,rapp}.

The initially produced quarkonia suffer from also cold nuclear matter effects before the formation of the hot medium. There are mainly three kinds of cold nuclear matter effects. 1) The shadowing effect~\cite{shadowing}: the parton distribution function in nuclei is different from the one in a free nucleon, and the quarkonium yield in heavy ion collisions is not a simple superposition of p+p collisions. 2) The Cronin effect~\cite{cronin1,cronin2}: before two gluons fuse into a quarkonium, the gluons obtain momentum from multiple scattering with surrounded nucleons. This extra momentum will be inherited by the quarkonium and leads to a transverse momentum broadening. 3) The nuclear absorption~\cite{gerschel1}: the multiple scattering between the quarkonium (or its pre-resonance state) and spectator nucleons leads to a normal suppression of quarkonia. Experimentally, the cold nuclear matter effects can be extracted from p+A collisions where the hot medium is not expected.

The nuclear matter effects on quarkonium production in heavy ion collisions can be described by the so-called nuclear modification factor $R_{AA}=N_{AA}/\left(N_{bin} N_{pp}\right)$, where $N_{bin}$ denotes the number of nucleon+nucleon
collisions for a given centrality, and $N_{pp}$ and $N_{AA}$ are respectively the quarkonium yields
integrated over momentum in p+p and A+A collisions. The color screening leads to $R_{AA}<1$ and the regeneration cancels partly the suppression. Since the colliding energy of heavy ion collisions at the Super Proton Synchrotron (SPS), RHIC and LHC increases by two orders of magnitude, the produced fireball temperature increases from around $T_c$ at SPS to about (2-3)$T_c$ at LHC, the quarkonium yield is expected to behave very differently from SPS to LHC. However, from the experimental data, while there is a clear energy dependence of the momentum integrated $J/\psi$ $R_{AA}$ in central heavy ion collisions, its trend on collision centrality is similar at all energies and there is always $R_{AA}<1$~\cite{raa}.

The particle yield is a global quantity, the momentum integration smears the fireball structure and the cold and hot nuclear matter effects at different energies. To have a deep insight into what is happening to the quarkonium motion in the hot medium, one needs to concentrate on the quarkonium momentum distribution. Different from the longitudinal motion which inherits the initial colliding kinematics via momentum conservation, the transverse motion in heavy ion collisions is developed during the dynamical evolution of the system. The microscopically high particle density and multiple scatterings play essential role in the development of the finally observed transverse momentum distribution. The distribution is therefore sensitive to the medium properties like the equation of state. The study on the transverse motion has been well documented in light quark sectors at all
energies~\cite{rhicexp1,rhicexp2,rhicexp3,rhicexp4,hecke98}. For quarkonia, we expect that their transverse momentum distribution can help us to probe the detailed structure of the fireball and differentiate between the production and suppression mechanisms.

We will focus on the charmonium production in this paper, especially the $J/\psi$ production. The extension to bottonia and excited states are straightforward. We first introduce a detailed transport approach for quarkonium production in high energy nuclear collisions which incorporates a hydrodynamic description for the space-time evolution of the hot medium and a transport equation for the quarkonium motion in the medium. We second analyze the data on charm quark production in the initial state and its energy loss in the hot medium which are inputs for the charmonium regeneration. We then compare the model calculations in A+A collisions with the data, including the transverse momentum distribution $dN_{J/\psi}/p_Tdp_T$, the differential  nuclear modification factor $R_{AA}(p_T)$, and the second $p_T$ moment $\langle p_T^2\rangle_{AA}$. We summarize in the end.

\section{A Transport Approach for Charmonium Motion in Hot Medium}
\label{s2}
In order to extract information about the nature of the medium from
charmonium production in heavy ion collisions, the medium created in the initial stage and
the charmonia produced in the initial stage and in the medium should be treated both dynamically.

After the system created in a heavy ion collision reaches local equilibrium, the space-time evolution of the strongly interacting medium can be described by the ideal hydrodynamic equations
\begin{eqnarray}
\label{hydro}
&& \partial_\mu T^{\mu\nu}=0, \nonumber\\
&& \partial_\mu j^{\mu}=0,
\end{eqnarray}
where $T_{\mu\nu}=(\epsilon+p)u^{\mu}u^{\nu}-g^{\mu\nu}p$
is the energy-momentum tensor of the
medium, $j^{\mu}=nu^{\mu}$ the conserved baryon current,
$u^{\mu}$ the four-velocity of the fluid cell, $\epsilon$ the energy density,
$p$ the pressure, and $n$ the baryon density.
The solution of the hydrodynamic equations provides the local
temperature $T(x)$, baryon density $n(x)$ and fluid velocity $u_\mu(x)$ of the medium which will be used in the
charmonium suppression and regeneration rates. To simplify the numerical calculations, we employ
the well tested 2+1 dimensional version of the hydrodynamics, considering the Hubble-like
expansion and boost invariant initial condition for the longitudinal motion. Taking the proper time $\tau=\sqrt{t^2-z^2}$ and space-time rapidity $\eta=1/2\ln[(t+z)/(t-z)]$ instead of the time $t$ and longitudinal coordinate $z$, the above equations can be simplified as\cite{xiangzhuang}
\begin{eqnarray}
\label{hydro3}
&& \partial_{\tau}E+\nabla\cdot{\bf M} = -(E+p)/{\tau}, \nonumber\\
&& \partial_{\tau}M_x+\nabla\cdot(M_x{\bf v}) = -M_x/{\tau}-\partial_xp,\nonumber\\
&& \partial_{\tau}M_y+\nabla\cdot(M_y{\bf v}) = -M_y/{\tau}-\partial_yp,\nonumber\\
&& \partial_{\tau}R+\nabla\cdot(R{\bf v}) = -R/{\tau}
\end{eqnarray}
with the definitions $E=(\epsilon+p)\gamma^2-p$, ${\bf M}=(\epsilon+p)\gamma^2 {\bf v}$
and $R=\gamma n$, where ${\bf v}$ and $\gamma$ are the fluid velocity and Lorentz factor in the transverse plane.

To close the hydrodynamical equations one needs to know the equation
of state of the medium. We follow Ref.\cite{sollfrank} where the
deconfined phase at high temperature is an ideal gas of gluons and massless
$u$ and $d$ quarks plus 150 MeV massed $s$ quarks, and the
hadron phase at low temperature is an ideal gas of all known
hadrons and resonances with mass up to 2 GeV~\cite{pdg}. There is
a first order phase transition between these two phases. In the
mixed phase, the Maxwell construction is used. The mean field
repulsion parameter and the bag constant are chosen as $K$=450
MeV fm$^3$ and $B^{1/4}$=236 MeV to obtain the
critical temperature $T_c=165$ MeV~\cite{sollfrank} at vanishing baryon number
density.

For the initialization of the hot medium, we take the same treatment as in Ref.~\cite{xiangzhuang} for collisions
at SPS and RHIC and Ref.~\cite{hirano} at LHC. The maximum temperature of the medium at the starting time
$\tau_0=0.6$ fm/c is, for instance, $T_0=484$ and $430$ MeV corresponding respectively to the observed charge number
density $dN_{ch}/dy=1600$ and $1200$~\cite{chargealice} in the mid and forward
rapidity regions for central 2.76 TeV Pb+Pb collisions at LHC. For the decay of the fluid, we assume that the medium maintains chemical and thermal equilibrium until
the energy density of the system drops to a value of $60$ MeV/fm$^3$, when the
hadrons decouple and their momentum distributions are fixed.

We now turn to the charmonium motion in the hot medium. Since a charmonium is so heavy, its equilibrium with the medium can
hardly be reached, we use a transport approach to describe
its distribution
function $f_\Psi(x,{\bf p}|{\bf b})$ in the phase space
$(x,{\bf p})$ in heavy ion collisions with impact
parameter ${\bf b}$,
\begin{equation}
p^\mu \partial_\mu f_\Psi = - C_\Psi f_\Psi + D_\Psi,
\label{trans1}
\end{equation}
where the lose and gain terms $C_\Psi(x,{\bf p}|{\bf
b})$ and $D_\Psi(x,{\bf p}|{\bf b})$ come from the charmonium dissociation and regeneration. We have neglected here the elastic scattering, since
the charmonium mass is much larger than the typical medium temperature. Considering the contribution from the feed-down of the excited
states to the finally observed ground state~\cite{decay}, we should consider the transport equations for $\Psi=J/\psi,\ \psi'$ and $\chi_c$, when we
calculate the $J/\psi$ distribution $f_{J/\psi}$.

Taking the proper time $\tau$, space-time rapidity $\eta$, the momentum rapidity
$y=1/2\ln\left[(E+p_z)/(E-p_z)\right]$ and transverse energy
$E_t=\sqrt {E^2-p_z^2}$ to replace $t$, $z$, $p_z$ and $E=\sqrt{m^2+{\bf
p}^2}$, the transport equation can be rewritten as
\begin{eqnarray}
\left[\cosh(y-\eta)\partial\tau+{\sinh(y-\eta)\over \tau}\partial
\eta+{\bf v}_t\cdot\nabla_t\right]f_\Psi=- \alpha_\Psi f_\Psi+\beta_\Psi
\label{trans2}
\end{eqnarray}
with the dissociation and regeneration rates $\alpha_\Psi(x,{\bf
p}|{\bf b}) = C_\Psi(x,{\bf p}|{\bf b})/E_t$ and $\beta_\Psi(x,{\bf p}|{\bf b}) = D_\Psi(x,{\bf p}|{\bf b})/E_t$, where the third term in the square bracket arises from the free streaming of $\Psi$ with transverse velocity ${\bf v}_T={\bf p}_T/E_T$ which leads to a strong leakage effect at SPS
energy~\cite{spsleakage}.

Color screening is an ideal and static description for the charmonium dissociation in hot medium. To dynamically treat the suppression process, we should consider the charmonium interaction with the ingredients of the medium. At high temperature the gluon
dissociation $\Psi + g \to c+\bar c$ is the dominant suppression process, and the corresponding dissociation rate $\alpha$ can be expressed as
\begin{equation}
\label{loss}
\alpha_\Psi=\frac{1}{2E_T}\int{d^3{\bf k}\over (2\pi)^3
2E_g}\sigma_{g\Psi}({\bf p},{\bf k},T)4F_{g\Psi}({\bf p},{\bf k})f_g({\bf k},T,u_\mu),
\end{equation}
where $E_g$ is the gluon energy, $F_{g\Psi}=\sqrt{(p\cdot k)^2-m_\Psi^2m_g^2}=p\cdot k$ the flux factor, and $f_g$ the gluon thermal distribution as a function of the local temperature $T(x|{\bf b})$ and fluid velocity $u_\mu(x|{\bf b})$ of the medium determined by the hydrodynamics (At RHIC and LHC energy one can safely neglect the baryon density). The dissociation cross section in vacuum $\sigma_{g\Psi}({\bf p},{\bf k},0)$ can be derived through the operator production expansion (OPE) method with a perturbative Coulomb wave function~\cite{ope1,ope2,ope3,ope4}. However, the method is no longer valid for loosely bound states at high temperature. To reasonably describe the temperature dependence of the cross section, we consider the geometric relation between the average charmonium size and the cross section,
\begin{equation}
\label{crosssection}
\sigma_{g\Psi}({\bf p},{\bf k},T)={\langle r^2\rangle_\Psi(T)\over \langle r^2\rangle_\Psi(0)}\sigma_{g\Psi}({\bf p},{\bf k},0).
\end{equation}
The average distance $\langle r^2\rangle_\Psi(T)$ between the $c$ and $\bar c$ is calculated via potential model~\cite{pot1,pot2} with lattice simulated heavy quark potential~\cite{petreczky} at finite temperature. When $T$ approaches to the dissociation temperature $T_d$, the distance and in turn the cross section go to infinity which means a full charmonium dissociation. Note that the dissociation here does not happen suddenly at $T_d$ but a continuous process in the temperature region $T\leq T_d$.

The regeneration rate $\beta$ is connected to the dissociation rate $\alpha$ via
the detailed balance between the gluon dissociation process $g+\Psi\to c+\bar c$ and its inverse
process $c+\bar c\to g+\Psi$~\cite{thews,yan}. To take into account the
relativistic effect on the dissociation cross section which is derived in non-relativistic limit and to avoid non-physics divergence in the
regeneration cross section, we should replace
the charmonium binding energy by the gluon threshold energy in the calculations of $\alpha$ and $\beta$~\cite{apo04}.

Different from the gluons and light quarks which are ingredients of the hot medium, while heavy quarks produced via initial hard processes will lose energy when they pass through the hot medium, they may not be fully thermalized, since they are so heavy. The heavy quark distribution functions $f_Q$ and $f_{\bar Q}$ which appear in the regeneration rate $\beta$ are in principle between the pQCD and equilibrium distributions. From the experimental data at RHIC and LHC, the observed
large quench factor~\cite{rhicdquench,dquench} and elliptic flow~\cite{rhicdv2,dv2}
for charmed mesons indicate that the charm quarks interact strongly with the
medium. Therefore, one can take, as a good approximation, a
kinetically thermalized phase space distribution $f_c$ for charm quarks. We also take the approximation for bottom quarks in the calculation of charmonium distributions at LHC energy, where  the B decay contribution to the inclusive charmonia should be considered.
Neglecting the creation and annihilation for charm quark and antiquark pairs
inside the medium, the spacial charm quark (antiquark) density
number $\rho_c(x|{\bf b})=\int d^3{\bf q}/(2\pi)^3f_c(x,{\bf q}|{\bf b})$ satisfies the conservation law
\begin{equation}
\label{cflow}
\partial_\mu\left(\rho_c u^\mu\right)=0
\end{equation}
with the initial density determined by the nuclear geometry
\begin{equation}
\label{rhoc}
\rho_c(x_0|{\bf b})=\frac{T_A({\bf x}_t)T_B({\bf x}_t-{\bf b})\cosh\eta}
{\tau_0} {d\sigma_{pp}^{c\bar c}\over d\eta},
\end{equation}
where $T_{A,B}({\bf x}_t)=\int_{-\infty}^{+\infty}\rho_{A,B}(\vec{r}) dz$ are the thickness
functions at transverse
coordinate ${\bf x}_t$ defined in the Glauber model~\cite{glauber} with a Woods-Saxon
distribution for nucleon density in nuclei A and B, and
$d\sigma_{pp}^{c\bar c}/d\eta$ is the rapidity distribution of charm
quark production cross section in p+p collisions.

In the hadron phase of the fireball with temperature $T<T_c$, there are many effective
models that can be used to calculate the inelastic cross sections between charmonia and
hadrons\cite{hadronphase}. For $J/\psi$ the dissociation cross section is about a few mb
which is comparable with the gluon dissociation cross section. Considering that the hadron
matter appears later in the evolution of the fireball when the ingredient density of the system
is much lower in comparison with the early hot and dense period, we neglect the
charmonium production and suppression in hadron gas.

The transport equation can be solved analytically
with the explicit solution
\begin{eqnarray}
\label{solution}
f_\Psi\left({\bf p}_T,y,{\bf
x}_T,\eta,\tau\right)&=&f_\Psi\left({\bf p}_T,y,{\bf
X}(\tau_0),H(\tau_0),\tau_0\right)\times\nonumber\\
&&e^{-\int^{\tau}_{\tau_0}{d\tau'\over \Delta(\tau')}
\alpha_\Psi\left({\bf p}_T,y,{\bf X}(\tau'),H(\tau'),\tau'\right)}\nonumber\\
&+&\int^{\tau}_{\tau_0}{d\tau'\over \Delta(\tau')} \beta_\Psi\left({\bf
p}_t,y,{\bf
X}(\tau'),H(\tau'),\tau'\right)\times\nonumber\\
&&e^{-\int^{\tau}_{\tau'}{d\tau''\over \Delta(\tau'')}\alpha_\Psi\left({\bf
p}_t,y,{\bf
X}(\tau''),H(\tau''),\tau''\right)}
\end{eqnarray}
with
\begin{eqnarray}
\label{xh}
&& {\bf X}(\tau')={\bf x}_T-{\bf
v}_T\left[\tau\cosh(y-\eta)
-\tau'\Delta(\tau')\right],\nonumber\\
&& H(\tau')=y-\arcsin\left(\tau/\tau' \sinh(y-\eta)\right),\nonumber\\
&&
\Delta(\tau')=\sqrt{1+(\tau/\tau')^2 \sinh^2(y-\eta)}.
\end{eqnarray}
The first and second terms on the right-hand side of the solution
(\ref{solution}) indicate the contributions from the initial
production and continuous regeneration, respectively, and both
suffer from the gluon dissociation in the medium. Since the regeneration
happens in the deconfied phase, the regenerated quarkonia
would have probability to be dissociated again by the surrounding gluons.
The coordinate shifts ${\bf x}_T \to {\bf
X}$ and $\eta \to H$ in the solution (\ref{solution})
reflect the leakage effect in the transverse and longitudinal
directions.

Suppose the charmonium formation time and the collision time for the two nuclei to pass through to each other at RHIC and LHC energy are less
than the formation time $\tau_0$ of the hot medium, all the cold nuclear matter effects on the initially
produced charmonia would cease before the QGP evolution. Therefore, they can be reflected in the
initial charmonium distribution $f_\Psi$ at time $\tau_0$. Based on a model dependent approach~\cite{lourenco}, the absorption cross section of $J/\psi$ decreases as a function of colliding energy. At RHIC energy the cross section reduces to about 3 mb and should be even smaller at LHC energy. In our approach, the effect of absorption by the cold nuclear matter is ignored. We take into account the nuclear shadowing and Cronin effects at both RHIC and LHC energies. In this case the initial distribution of the transport equation (\ref{trans2}) can be obtained from a geometrical superposition of p+p collisions, along with the modifications due to the shadowing and Cronin effects. As far as the cold nuclear matter concern, the attenuation of the $c\bar c$ dipole may offer an alternative explanation to the observed nuclear absorption~\cite{kopeliovich} in p+A collisions. While the nuclear absorption reduces the initial yield, the final production, in the limit of high-energy nuclear collisions, is dominated by total number of charm quarks and the later stage interactions near freeze-out, see discussions in the following sections.

The Cronin effect broadens the momentum distribution of the initially produced charmonia in heavy ion collisions. In pA and/or AA collisions, prior to any hard
scatterings, the incoming partons (both gluons and quarks) experience multiple scatterings via soft gluon exchanges.  The initial scatterings lead to additional
transverse momentum broadening which is inherited by produced particles including the charmonia~\cite{esumi97}. The Cronin effect is caused
by soft interactions and rigorous calculations for the effect are not available. However, the effect is often treated as random motion.
Inspired from a random-walk picture, we take a Gaussian smearing~\cite{gaus,liub} for the modified transverse momentum distribution

\begin{equation}
\label{cronin}
\overline
f^{pp}_\Psi({\bf x},{\bf p},z_A,z_B|{\bf b})={1\over \pi a_{gN} l} \int
d^2{\bf p}_T' e^{-{\bf p}_T^{'2}\over a_{gN} l}f^{pp}_\Psi(|{\bf
p}_T-{\bf p}_T'|,p_z),
\end{equation}
where
\begin{equation}
l({\bf x},z_A,z_B|{\bf b})=\frac{1}{\rho} \left(\int_{-\infty}^{z_A}\rho(z,{\bf x_T}) dz + \int_{z_B}^{+\infty}\rho(z,{\bf x_T-b}) dz\right)
\label{ll}
\end{equation}
is the path length of the initial gluons in nuclei before fusing into a charmonium at ${\bf x}$, $z_A$ and $z_B$ are
the longitudinal coordinates of the two nucleons where the two gluons come from, $a_{gN}$ is the averaged charmonium squared transverse
momentum gained from the gluon scattering with a unit of length of nucleons,
and $f^{pp}_\Psi({\bf p})$ is the charmonium momentum distribution in a nuclear medium free p+p collision. The Cronin parameter
$a_{gN}$ is usually extracted from corresponding p+A collisions where the cold nuclear matter effects are dominant.
The transverse momentum distributions are needed in order to fix the value of $a_{gN}$ experimentally. Considering the absence of p+A
collision data at $\sqrt{s_{NN}}=2.76$ TeV, we take $a_{gN}=0.15$ (GeV/c)$^2$/fm as suggested in Ref.~\cite{emp,emp1,thews}. For collisions at
SPS ($\sqrt{s_{NN}} \sim 20$ GeV) and  RHIC ($\sqrt{s_{NN}}=200$ GeV) we take the values of $a_{gN}=0.075$~\cite{xiangzhuang}  and
0.1~\cite{liu1} (GeV/c)$^2$/fm, respectively.

Assuming that the emitted gluon in the gluon fusion process $g+g\to
\Psi+g$ is soft in comparison with the initial gluons and the
produced charmonium and can be neglected in kinematics,
corresponding to the picture of color evaporation model at
leading order~\cite{Fritzsch:1977ay,cem1,cem2}, the longitudinal
momentum fractions of the two initial gluons are
calculated from the momentum conservation,
\begin{equation}
\label{x}
x_{1,2}={\sqrt{m_\Psi^2+p_T^2}\over \sqrt{s_{NN}}} e^{\pm y}.
\end{equation}
The free distribution
$f_\Psi^{pp}({\bf p})$ can
be obtained by integrating the elementary partonic processes,
\begin{equation}
\label{fg}
{d\sigma_\Psi^{pp}\over dp_Tdy}= \int dy_g x_1 x_2 f_g(x_1,\mu_F)
f_g(x_2,\mu_F) {d\sigma_{gg\to\Psi g}\over d\hat t},
\end{equation}
where $f_g(x,\mu_F)$ is the gluon distribution in a free proton, $y_g$ is the emitted
gluon rapidity, $d\sigma_{gg\to\Psi g}/ d\hat t$ is
the charmonium momentum distribution produced from a gluon fusion
process, and $\mu_F$ is the factorization scale of the fusion process.

Now we consider
the shadowing effect. The distribution function $\overline
f_i(x,\mu_F)$ for parton $i$ in a nucleus differs from a superposition of the
distribution $f_i(x,\mu_F)$ in a free nucleon. The nuclear shadowing
can be described by the modification factor $R_i=\overline f_i/(Af_i)$.
To account for the spatial dependence of the shadowing in a finite
nucleus, one assumes that the inhomogeneous shadowing is
proportional to the parton path length through the nucleus~\cite{shadpath},
which amounts to consider the coherent interaction of the incident
parton with all the target partons along its path length. Therefore,
we replace the homogeneous modification factor $R_i(x,\mu_F)$ by an
inhomogeneous one~\cite{vogtshad}
\begin{equation}
 {\cal R}_i(x,\mu_F,{\bf x})=1+A\left(R_i(x,\mu_F)-1\right)T_A({\bf x}_T)/T_{AB}(0)
\end{equation}
with the definition $T_{AB}({\bf b})=\int d^2{\bf x}_T T_A({\bf x}_T) T_B({\bf x}_T-{\bf b})$.
We employ in the following the EKS98 package~\cite{eks98} to evaluate the homogeneous
ratio $R_i$, and the factorization scale is taken as
$\mu_F=\sqrt{m_\Psi^2+p_T^2}$.

Replacing the free distribution $f_g$ in (\ref{fg}) by the modified
distribution $\overline f_g=Af_g{\cal R}_g$ and then taking into account the Cronin
effect (\ref{cronin}), we finally get the initial charmonium distribution for the
transport equation (\ref{trans2}),
\begin{eqnarray}
\label{initial}
f_\Psi({\bf x},{\bf p},\tau_0|{\bf b})&=&{(2\pi)^3\over E_T\tau_0}\int dz_Adz_B\rho_A({\bf x}_T,z_A)\rho_B({\bf x}_T,z_B)\nonumber\\
&\times&{\cal R}_g(x_1,\mu_F,{\bf x}_T){\cal R}_g(x_2,\mu_F,{\bf x}_T-{\bf b})\overline f_\Psi^{pp}({\bf x},{\bf p},z_A,z_B|{\bf b}).
\end{eqnarray}
Now the only thing left is the distribution $f_\Psi^{pp}$ in a free p+p collision which
can be fixed by experimental data or by some model simulations.

\section{In-medium Heavy Quarks}
\label{s3}

We now discuss, from the point of view of the experimental results, the heavy quark production in the initial stage and the interaction between heavy quarks and the hot medium. The former controls the fraction of the regeneration in the total quarkonium yield, and the latter characterizes the transverse momentum properties of the regenerated quarkonia. Both are reflected in the heavy quark distributions $f_Q$ and $f_{\bar Q}$ in the regeneration rate $\beta$.

\begin{figure}[ht]
\centering
\includegraphics[width=0.55\textwidth]{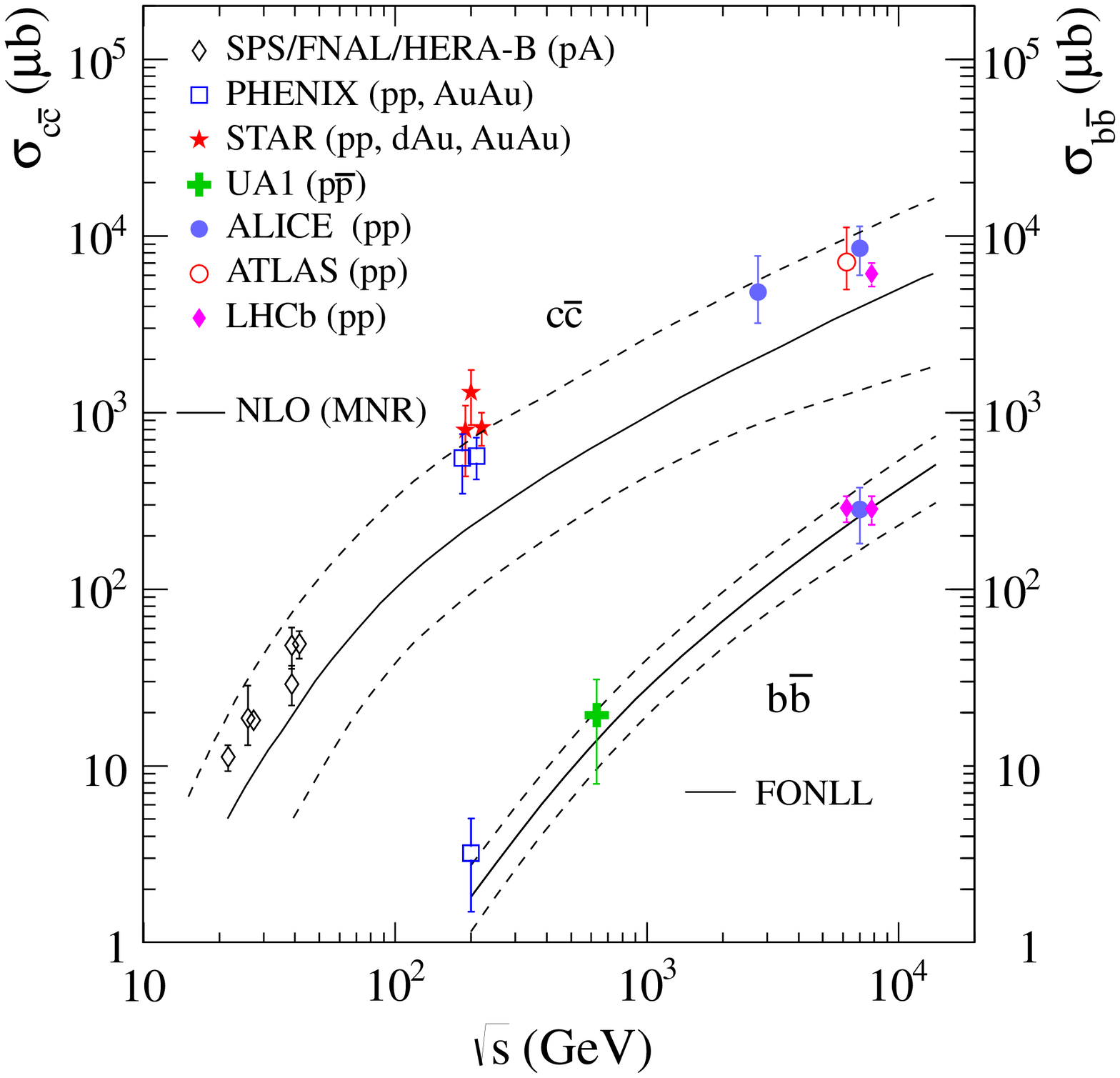}
\caption{(Color online) The total charm and bottom quark production cross sections as functions of colliding energy. The data are from the
NA16, NA27, E653, E743, E769, HERA-B, PHENIX, STAR, UA1, ALICE, ATLAS and LHCb
collaborations~\cite{cc_bb_fixed_target, cc_NA16, cc_NA27, cc_E653, cc_E743, cc_E769, cc_HERAB, cc_PHENIX_pp_AuAu, cc_STAR_pp, cc_STAR_dAu, cc_STAR_AuAu, cc_ALICE, cc_ATLAS, cc_LHCb, bb_PHENIX_pp, bb_UA1, bb_ALICE, bb_LHCb_2010, bb_LHCb_2011}. Some of the data points at 200 GeV and 7 TeV are shifted slightly along the horizontal axis for clarification. The solid lines show the NLO (MNR)~\cite{cc_NLO_NMR} and FONLL~\cite{heavy_FONLL} calculations, and the dashed lines depict the corresponding uncertainties.}
\label{fig1}
\end{figure}

Fig.\ref{fig1} shows the colliding energy dependence of the total charm and bottom
quark production cross sections $\sigma_{Q\bar Q}$ per p+p collision and the comparison with model calculations. For p+A and A+A collisions, the measured cross sections are scaled by the number of binary collisions $N_{bin}$.

The number $N_{Q\bar Q}$ of produced heavy quark pairs in A+A collisions at a
given centrality can be estimated from the heavy quark cross section together with the measurement of p+p inelastic cross section $\sigma^{in}_{pp}$ and the Glauber model
calculation of $N_{bin}$,
\begin{equation}
N_{Q\bar Q}=\sigma_{Q\bar Q}N_{bin}/\sigma^{in}_{NN}.
\end{equation}
Both $\sigma^{in}_{NN}$ and $N_{bin}$ increase with colliding energy, but the
nuclear overlap function $T_{AA}=N_{bin}/\sigma^{in}_{NN}$ is energy independent. It represents the effective nucleon luminosity in the A+A collision process.
Tab.\ref{tab1} shows $T_{AA}$ in Au+Au and Pb+Pb collisions at different centrality bins.
\begin{table}
\centering
\caption{The nuclear overlap function $T_{AA}$ in Au+Au and Pb+Pb collisions at five centrality bins, based on the numbers from Refs.~\cite{TAA_RHIC, TAA_LHC}. The unit is $mb^{-1}$.}
\vspace{0.2cm}
\begin{tabular}{cccccc}
\hline
System&0-10\%&10-20\%&20-40\%&40-60\%&60-80\%\\
\hline
Au+Au&22$\pm$1.3 &14$\pm$0.88 &7.1$\pm$0.50 &2.4$\pm$0.27 &0.58$\pm$0.11\\
Pb+Pb&23$\pm$2.1 &14$\pm$2.2 &6.8$\pm$2.3&2.0$\pm$0.92 &0.42$\pm$0.29\\
\hline
\end{tabular}\label{tab1}
\end{table}
\begin{table}
\centering
\caption{The number of $c\bar c$ pairs produced in A+A collisions with different centrality bins at SPS, RHIC and LHC energies. The charm production cross section at 17.2 GeV and 5.5 TeV are extrapolated from the measurements at 21.6 GeV and 7 TeV, respectively, based on the upper boundary of the NLO (MNR) calculations~\cite{cc_NLO_NMR}.}
\vspace{0.2cm}
\begin{tabular}{cccccc}
\hline
System&0-10\%&10-20\%&20-40\%&40-60\%&60-80\%\\
\hline
17.2 GeV Pb+Pb&0.13$\pm$0.03&0.081$\pm$0.019&0.039$\pm$0.015&0.012$\pm$0.006& 0.0024$\pm$0.0017\\
200 GeV Au+Au &18$\pm$4 &11$\pm$2 & 5.7$\pm$1.3 & 2.4$\pm$0.5& 0.47$\pm$0.13\\
2.76 TeV Pb+Pb&110$\pm$65 &67$\pm$40 &33$\pm$22&10$\pm$7 &2.0$\pm$1.8\\
5.5~ TeV Pb+Pb &142$\pm$35 &87$\pm$24 &42$\pm$17&12$\pm$6 &2.6$\pm$1.9\\
\hline
\end{tabular}\label{tab2}
\end{table}
\begin{table}
\centering
\caption{The number of $b\bar b$ pairs produced in A+A collisions with different centrality bins at RHIC and LHC energies. The bottom production cross section at 2.76 TeV and 5.5 TeV are extrapolated from the measurement at 7 TeV based on the FONLL calculations~\cite{heavy_FONLL}.}
\vspace{0.2cm}
\begin{tabular}{cccccc}
\hline
System&0-10\%&10-20\%&20-40\%&40-60\%&60-80\%\\
\hline
200 GeV Au+Au&0.07$\pm$0.04&0.045$\pm$0.026 & 0.023$\pm$0.013 & 0.0077$\pm$0.0045& 0.0019$\pm$0.0011\\
2.76 TeV Pb+Pb&2.3$\pm$0.4 & 1.5$\pm$0.3 &0.70$\pm$0.26 & 0.21$\pm$0.10 & 0.044$\pm$0.030\\
5.5~ TeV Pb+Pb &4.9$\pm$0.9 & 3.1$\pm$0.6 & 1.5$\pm$0.5 & 0.44$\pm$0.21 & 0.094$\pm$0.062\\
\hline
\end{tabular}\label{tab3}
\end{table}

The produced numbers of $c\bar c$ and $b\bar b$ pairs in A+A collisions are listed in Tabs.\ref{tab2} and \ref{tab3} for five centrality bins at SPS, RHIC and LHC energies.  It is assumed that there are no thermal production and annihilation of heavy quarks in the hot medium. For central collisions, the number of $c\bar c$ pairs
is much less than one at SPS, but increases rapidly to around 20 at RHIC and even becomes
more than 100 at LHC. Since the nuclear geometry for Pb+Pb is almost the same as for Au+Au, see Tab.\ref{tab1}, the rapid change in $N_{c\bar c}$ is from the energy dependence of the cross section shown in Fig.\ref{fig1}.  We can schematically estimate the degree of regeneration at different energies, by considering the relation between the number $N_\Psi^{reg}$ of regenerated charmonia and the number $N_{c\bar c}$ of charm quark pairs for a homogeneous fireball $N_\Psi^{reg}\sim N_{c\bar c}^2$. It is clear that the charmonium regeneration is negligible at SPS, but starts to play an important
role at RHIC and becomes dominant at LHC. For the number of $b\bar b$
pairs in central collisions, it is much less than one at RHIC and smaller than
the number of $c\bar c$ pairs at SPS, which guarantees the simplification to neglect bottomonium
regeneration at SPS and RHIC. However, the production of $b\bar b$ pairs in central collisions at LHC is similar to the production of $c\bar c$ pairs in semi-central collisions at RHIC, the bottomonium regeneration starts to play a role at LHC.

The above description on heavy quark production in A+A collisions is simply a superposition of p+p collisions and all the cold and hot nuclear matter effects have been neglected. While the nuclear shadowing at SPS and RHIC may not be so important, the gluon longitudinal momentum fraction $x$ defined in (\ref{x}) is very small at LHC and located in the strong shadowing region~\cite{shadowing}. The shadowing effect reduces the numbers
of charm and bottom quarks. Estimated from the centrality averaged EKS98
evolution~\cite{eks98}, we take a $20\%$ reduction for the charm and
bottom quark production cross sections in our calculations at LHC energy. Note that a $20\%$ reduction for the heavy quark cross section leads to a reduction of about $[1-(80\%)^2]=36\%$ for the yield of regenerated quarkonia.

\begin{figure}[ht]
\centering
\includegraphics[width=0.7\textwidth]{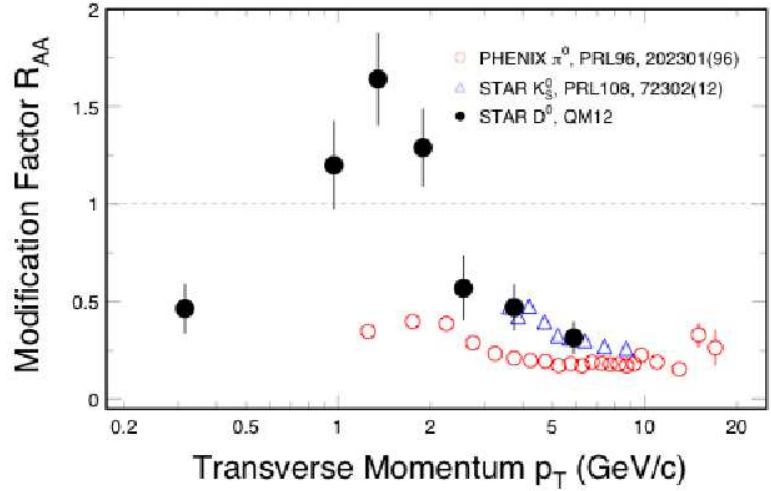}
\caption{(Color online) The differential nuclear modification factor $R_{AA}$ as a function of transverse momentum for mesons $\pi$, $K$ and $D$ produced in central Au+Au collisions at $\sqrt {s_{NN}}=200$ GeV. The data are from Refs.~\cite{pion0,kaons,d0meson}}
\label{fig2}
\end{figure}

While the thermal production and annihilation of heavy quarks can be reasonably neglected, their momentum distribution will be affected by the hot medium. When the heavy quarks pass through the medium, they lose energy via elastic scattering and gluon radiation~\cite{shanshancao}, and the initial pQCD distribution is gradually changed and approaches to the thermal distribution. Fig.\ref{fig2} shows the experimental data on the nuclear modification factor $R_{AA}$ as a function of transverse momentum $p_T$ for mesons $\pi$, $K$ and $D$ produced in central Au+Au collisions at RHIC energy. In the low $p_T$ region with $p_T<5$ GeV/c, $D$ is dramatically different from the other light mesons. Around $p_T\sim 1.5$ GeV/c, there is a peak clearly above unit for $D$ mesons. Since the behavior of $D$ mesons is characterized by the constituent charm quarks, the strong mass dependence of the nuclear modification factor comes from the strong interaction between charm quarks and the medium. The energy loss of charm
quarks
shifts $D$ mesons from high $p_T$ to intermediate $p_T$, and the collective flow of the medium shifts $D$ mesons from low $p_T$ to intermediate $p_T$. These two effects, especially the collective motion of charm quarks originated from the interaction with the medium, lead to the peak of $R_{AA}$ for $D$ mesons~\cite{PBG,MinHe}.

Taking together with the observed sizable elliptic flow $v_2$ of none photonic electrons, measured by both PHENIX~cite{chicdv2} and STAR~\cite{stardv2} experiments,  one can reasonably conclude that significant amount of  charm quarks are thermalized in nuclear collisions at RHIC energies. At LHC energy, much stronger effect of energy loss and D-meson $v_2$ have been observed for both open charms and open bottoms~\cite{dv2}, one can take thermal distributions for charm quarks safely and bottom quarks reasonably.

\section{Charmonium Transverse Momentum Distribution}
\label{s4}
\begin{figure}[ht]
\centering
\includegraphics[width=0.55\textwidth]{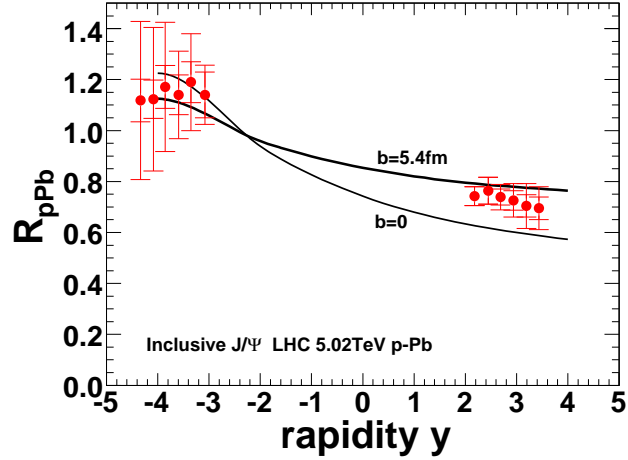}
\caption{(Color online) The transverse momentum integrated nuclear modification factor $R_{pA}$ as a function of rapidity for $J/\psi$ in p+Pb collisions at LHC energy $\sqrt {s_{NN}}=5.02$ TeV. The minimum bias data are from \cite{alicepa}, and the two lines are the model calculations with impact parameter $b=0$ and $5.4$ fm. }
\label{fig3}
\end{figure}

For p+A collisions, it is expected that there is almost no possibility to form hot medium during the evolution of the collisions. Therefore, the parameters for the cold nuclear matter effects can be experimentally extracted from p+A collisions. Neglecting the dissociation and regeneration, $\alpha_\Psi=\beta_\Psi=0$, the final charmonium distribution in the phase space is just the initial distribution (\ref{initial}) of the transport equation (\ref{trans2}). Integrating the phase-space distribution over the target nucleus where the cold nuclear matter effects happen, one obtains the charmonium momentum distribution. Fig.\ref{fig3} shows the $p_T$ integrated $R_{pA}$ as a function of rapidity for $J/\psi$ in p+Pb collisions at LHC energy $\sqrt {s_{NN}}=5.02$ TeV. Since the forward and backward rapidity are located in different shadowing regions, $J/\psi$s are suppressed in the forward rapidity due to nuclear shadowing and enhanced in the backward rapidity due to the nuclear anti-shadowing~\cite{antishadowing}. The model
calculation with the code EKS98~\cite{eks98} to describe the shadowing effect is in good agreement with the minimum bias data, when we take impact parameter $b=5.4$ fm. For the most central collisions with $b=0$, the shadowing and anti-shadowing effects become the strongest.

\begin{figure}[ht]
\centering
\includegraphics[width=0.8\textwidth]{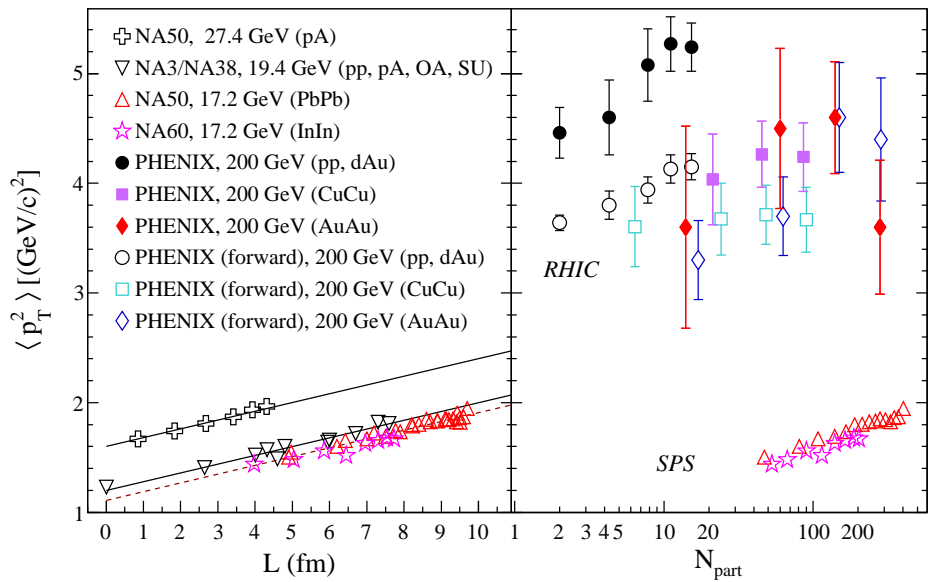}
\caption{(Color online) The mean squared transverse momentum $\langle p_T^2 \rangle$ of $J/\psi$ as a function of the averaged traveling length $L$ of the two gluons before their fusion into a charmonium (left panel) and as a function of the number $N_{part}$ of participant nucleons (right panel) in p+p, p+A and A+A collisions at mid ($|y|<0.35$) and forward ($1.2<|y|<2.2$ for A+A and $1.2<y<2.2$ for $d$+Au) rapidity at SPS and RHIC energy. The data are taken from
Refs.~\cite{pt2_NA3_pp_pA, pt2_NA38_pA_OA, pt2_NA38_SU, pt2_NA50_2000, pt2_NA50_1996, pt2_NA60, Jpsi_PHENIX_dAu2013, Jpsi_PHENIX_CuCu, Jpsi_PHENIX_AuAu2007}, and the three lines on the left panel indicate the linear relation (\ref{cronin}). For Cu+Cu and Au+Au collisions at RHIC, the covered transverse momentum range is $p_T \le 5$ GeV/c and the rest of the results are from all $p_T$ range. From the p+p collisions at RHIC, one finds that the values of $J/\psi$ $\langle p_T^2\rangle$ from different $p_T$ coverage are different no more than one standard deviation~\cite{Jpsi_PHENIX_dAu2013,Jpsi_PHENIX_dAule5}.
}
\label{fig4}
\end{figure}

We now turn to the mean squared transverse momentum $\langle p_{T}^2 \rangle$ of $J/\psi$. For p+A collisions, it is controlled by the initial Cronin effect~\cite{cronin1,cronin2} which results in a $p_T$ broadening in the final state. Integrating the smeared Gaussian distribution, see Eq. (\ref{cronin}), one obtains the path length dependence of the $\langle p_T^2\rangle$

\begin{equation}
\label{cronin5}
\langle p_T^2 \rangle = \langle p_T^2 \rangle_{pp}+a_{gN} L,
\end{equation}
where $\langle p_T^2\rangle_{pp}$ is the mean squared transverse momentum from p+p collisions. The left panel of Fig.~\ref{fig4} shows the results of $\langle p_T^2\rangle$ as a function of $L$ for p+p, p+A and A+A collisions at colliding energies $\sqrt {s_{NN}}$= 17.2, 19.4 and 27.4 GeV. The data for p+A collisions can well be described by the linear relation, Eq.~\ref{cronin5}, represented by the solid lines with the same slop parameter $a_{gN}=0.08$ (GeV/c)$^2$/fm. As one can see in the plot, the  values of  $\langle p_T^2\rangle_{pp}$ shows a clear energy dependence, namely, the higher the collision energy the larger the value: $\langle p_T^2\rangle_{pp}=$ 1.2 and 1.6 (GeV/c)$^2$ for $\sqrt{s_{NN}}=19.4$ and 27.4 GeV, respectively. The linear relation still holds for peripheral and central In+In collisions (open stars) and even peripheral and semi-central Pb+Pb collisions (open triangles) at $\sqrt{s_{NN}}$=17.2 GeV, see the dashed line in Fig.~\ref{fig4} left plot. However, once hot nuclear medium is formed in central heavy ion collisions one would expect the breakdown of the linear dependence.


\begin{figure}[ht]
\centering
\includegraphics[width=0.5\textwidth]{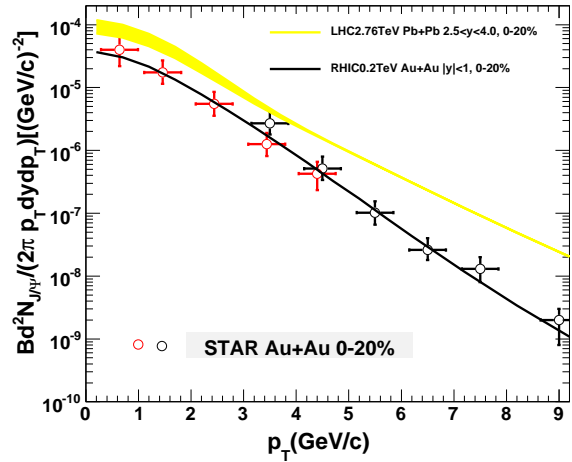}
\caption{(Color online) The $J/\psi$ transverse momentum distribution in central Au+Au collisions at RHIC and Pb+Pb collisions at LHC. The data are from Refs.\cite{starhighpt} and \cite{starlowpt}, the line is the model calculation at RHIC, and the band at LHC is due to the uncertainty in the charm quark cross section 0.4 mb$<d\sigma_{c\bar c}/dy<$0.5 mb. }
\label{fig5}
\end{figure}

Before we discuss the hot medium effects on $J/\psi$ $\langle p_T^2\rangle$ in heavy ion collisions at RHIC energy, shown in the right panel of Fig.\ref{fig4}, we need to first calculate the $J/\psi$ transverse momentum distribution itself. Integrating the known charmonium phase-space distribution $f_\Psi({\bf p}_T, y, {\bf x}_T, \eta,\tau)$, see the solution (\ref{solution}) of the transport equation (\ref{trans2}), over the hadronization hypersurface at time $\tau_c({\bf x}_T,\eta)$ determined by $T({\bf x}_T,\eta,\tau_c)=T_c$, one obtains, by using the Cooper-Frye formula~\cite{cooper}, the charmonium rapidity and transverse momentum distribution~\cite{xiangzhuang}
\begin{equation}
{d^2N_\Psi\over 2\pi p_Tdp_Tdy}={1\over (2\pi)^3}\int d^2{\bf x}_Td\eta\tau_cm_T\cosh(y-\eta)f_\Psi({\bf p}_T, y, {\bf x}_T, \eta,\tau_c),
\end{equation}
where $m_T=\sqrt{m^2+{\bf p}_T^2}$ is the $J/\psi$ transverse mass. Fig.\ref{fig5} shows the $J/\psi$ transverse momentum distribution in central Au+Au collisions at mid rapidity at RHIC energy and central Pb+Pb collisions at forward rapidity at LHC energy. The model calculation agrees well with the RHIC data, and the band at LHC is due to the uncertainty in the charm quark production cross section~\cite{longpaper}, 0.4 mb $<d\sigma_{c\bar c}/dy <$ 0.5 mb, shown in Fig.\ref{fig1}. At the moment there are no LHC data for the $p_T$ distribution. Since most of the initially produced low $p_T$ $J/\psi$s which are sensitive to the medium are dissociated by thermal gluons, and the $J/\psi$ regeneration happens mainly in the low $p_T$ region due to the assumption of charm quark thermalization, the uncertainty in $\sigma_{c\bar c}$ which controls the degree of regeneration affects mainly the low $p_T$ region. That is the reason why the band is very narrow at high $p_T$ and becomes wide at low $p_T$.

\begin{figure}[ht]
\centering
\includegraphics[width=0.5\textwidth]{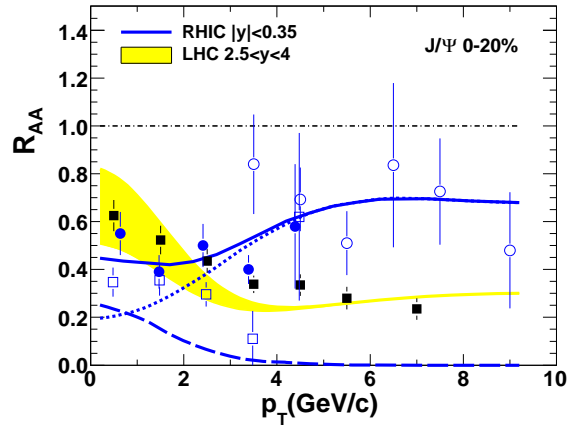}
\caption{(Color online) The $J/\psi$ nuclear modification factor as a function of transverse momentum in central Au+Au collisions at RHIC energy $\sqrt{s_{NN}}$=200 GeV and Pb+Pb collisions at LHC energy $\sqrt{s_{NN}}=2.76$ TeV. The data are from Refs.\cite{starhighpt,starlowpt} and \cite{aliceraapt}, and the lines and band are model calculations. The dotted, dashed and solid lines are the initial component, regeneration component and full result at RHIC, and the band at LHC is due to the uncertainty in the charm quark cross section 0.4 $<d\sigma_{c\bar c}/dy<$0.5 mb. }
\label{fig6}
\end{figure}

The transverse momentum distribution itself can hardly differentiate the hot mediums, its behavior at RHIC and LHC is similar.
To look into the nature of the medium, we should consider its comparison with the one in p+p collisions. To this end we calculate
the differential nuclear modification factor $R_{AA}(p_T)$ as a function of $p_T$~\cite{liu2}. The results for central Au+Au collisions at
RHIC and Pb+Pb collisions at LHC are shown and compared with the experimental data in Fig.~\ref{fig6}. The increase of $R_{AA}$
with only initial production, see the dotted line, comes from three aspects. One is the $p_T$ broadening via the Cronin
effect~\cite{cronin1,cronin2} happened in the initial stage. The second reason is the $p_T$ dependence of the gluon dissociation cross
section~\cite{thews,liu2}. Gluons with small energy are more likely to dissociate a $J/\psi$, or in other words, $J/\psi$s with low momentum
are easy to be eaten up by the hot medium. The last reason is the leakage effect with which the high momentum charmonia can escape the
suppression in the fireball. Note that the initial component becomes saturated at high $p_T$ due to the Gaussian smearing treatment (\ref{cronin})
of the Cronin effect. If we take instead the averaged linear relation (\ref{cronin5}) in the initial distribution (\ref{initial}), the initial component
will keep the increase at high $p_T$~\cite{liu2}. In comparison with the initially produced $J/\psi$s which carry high momentum from the hard
process, the regenerated $J/\psi$s from thermalized charm quarks that are mainly distributed at low momentum region, see clearly the dashed line
in Fig.~\ref{fig6}. In the low $p_T$ region, the competition between the initial production which increases with $p_T$ and the regeneration
which decreases with $p_T$ leads to a flat structure. While the initial production and regeneration are almost equally important in low $p_T$ region
and the latter even exceeds the former at extremely low $p_T$, the $J/\psi$ behavior at high $p_T$ is controlled by the initial component.
For collisions at LHC, there are obvious features arisen from the stronger suppression and regeneration. Comparing with the collisions at RHIC,
the fireball formed at LHC is much hotter, larger in size and lasts much longer. Therefore, the initially produced $J/\psi$s are all suppressed.
 At the same time the regeneration becomes dominant due to the large number of charm quark pairs shown in Table~\ref{tab2}. Since the
 initial production and regeneration dominate different $p_T$ regions, the stronger suppression leads to a lower $R_{AA}$ at high $p_T$ , and the
 stronger regeneration results in a higher $R_{AA}$ at low $p_T$ in comparison with the case at RHIC. Note that the uncertainty in the charm quark
 cross-section at LHC will only affect the value of $R_{AA}$ in the low $p_T$ region.

We now go back to the mean squared transverse momentum $\langle p_T^2\rangle$ versus the number of participant nucleons ($N_{part}$) in p+p, p+A and A+A collisions at SPS and RHIC energy. The results are shown in the right panel of Fig.~\ref{fig4}. Due to much higher collision energy, the values of $\langle p_T^2\rangle$ from RHIC are larger than that from collisions at SPS energies. As one can see, the strong centrality dependence in d+Au collisions does not show in either Au+Au or Cu+Cu collisions.
In the d+Au collisions at RHIC, if one converts the number of participant nucleons to the averaged path length ``L", the net increase in $\langle p_T^2\rangle$ from p+p to most central collisions is the same as in p+A collisions at the SPS energies. Since the averaged path length is similar in collisions at RHIC and SPS, this means a similar slope parameter $a_{gN}$. The resulting values of $a_{gN}$ are $ 0.11 \pm 0.05$ (GeV/c)$^2$/fm and $0.07 \pm 0.02$ (GeV/c)/fm$^2$ at mid-rapidity and forward-rapidity, respectively.  On the other hand, in both Au+Au or Cu+Cu collisions at RHIC, there is hardly any centrality dependence and, on average, the mean values of  $\langle p_T^2\rangle$ are lower than that from most central d+Au collisions, see Fig.~\ref{fig4}, right plot. These results indicate hot and dense medium formed in such heavy ion collisions. In the most central collisions at SPS, the Cronin induced increase in $\langle p_T^2\rangle$ is compensated by the hot medium color screening. At RHIC, both color screening and regeneration are important. Since regenerated $J/\psi$ tend to have lower transverse momentum which compensates the increase due to the Cronin effect.

\begin{figure}[ht]
\centering
\includegraphics[width=0.65\textwidth]{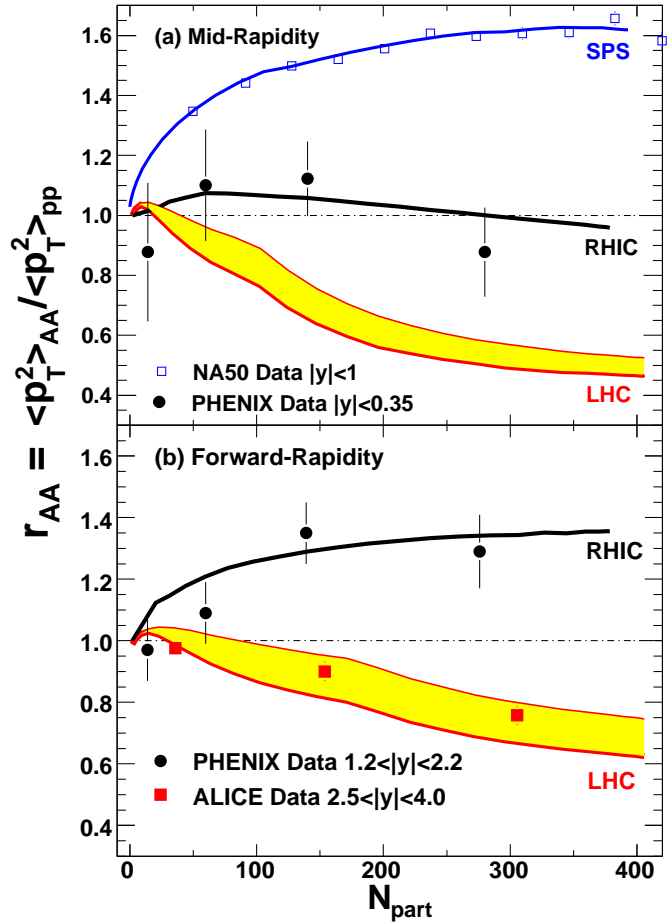}
\caption{(Color online) The $p_T$ nuclear modification factor $r_{AA}=\langle p_T^2\rangle_{AA}/\langle p_T^2\rangle_{pp}$ as a function of the number $N_{part}$ of participant nucleons for $J/\psi$ in Pb+Pb collisions at SPS and LHC energy $\sqrt{s_{NN}}=17.2$ GeV and 2.76 TeV and Au+Au collisions at RHIC energy $\sqrt{s_{NN}}=200$ GeV. The data are from Refs.\cite{na50,phenix,ptraaalice}, and the lines and bands are transport model calculations. The upper panel is for mid rapidity, and the lower panel is for forward rapidity. The bands at LHC are due to the uncertainty in the charm quark cross section 0.65 $<d\sigma_{c\bar c}/dy<$ 0.8 mb at mid-rapidity and 0.4 $<d\sigma_{c\bar c}/dy<$0.5 mb at forward-rapidity. }
\label{fig7}
\end{figure}

We often consider the ratio of particle distributions in A+A and p+p collisions to see clearly the nuclear matter effect. For instance, we calculate the yield nuclear modification factor $R_{AA}$ instead of the yield itself. Therefore, to extract the hot medium information from the $\langle p_T^2\rangle$ in A+A collisions, we introduce the $p_T$ nuclear modification factor~\cite{short},
\begin{equation}
\label{rpt}
r_{AA}=\frac{\langle p_T^2 \rangle_{AA}}{\langle p_T^2 \rangle_{pp}}.
\end{equation}
We will see clearly that the newly defined observable is really sensitive to the nature of the medium.
Here we are interested in the medium induced changes in $J/\psi$ transverse momentum distributions. Most of such medium interactions can be modeled as Brown motion such as the Cronin effect. The second moment is the lowest order to describe the distribution effectively. That is the reason we choose $\langle p_T^2 \rangle$ to characterize the $J/\psi$ transverse momentum distributions.  Fig.~\ref{fig7} shows the centrality dependence of the mid-rapidity $r_{AA}$ (upper plane) and forward-rapidity  (lower panel) in A+A collisions at SPS, RHIC and LHC energies.

Let us first focus on the mid rapidity. The SPS energy is low and only a small number of charm quarks are produced in the initial stage. Considering the expansion of the colliding system, the charm quark density at regeneration in the later stage is very small and the regeneration becomes negligible.  In this case, the Cronin effect, the leakage effect, and the $p_T$ dependence of the dissociation process lead to a broadening of the charmonium transverse momentum distribution. The $r_{AA}$ increases monotonously from unit in peripheral collisions to above 1.5 in most central collisions.

In comparison with SPS, the A+A collisions at LHC are the other limit where the regeneration becomes dominant. At LHC, the Cronin effect is still there, but the initially produced $J/\psi$s with enhanced transverse momentum are mostly dissociated by the large, hot and long-lived fireball, and only a small fraction can survive in the final state. The other hot nuclear matter effect, namely the regeneration turns to be the dominant source of the finally observed $J/\psi$s, due to the large number of charm quarks at LHC. Since thermalized charm quarks are distributed mostly in the low $p_T$ region, the regenerated $J/\psi$s carry low $p_T$. That is the reason why the $r_{AA}$ at LHC and SPS behaves in an opposite way: The $r_{AA}$ at LHC decreases monotonously from unit in peripheral collisions to around 0.5 in most central collisions.

The case at RHIC is in between SPS and LHC. Both the dissociation and regeneration are stronger than at SPS and weaker than at LHC. In this case, the low $p_T$ region is controlled by the equally important initial production and regeneration, see Fig.\ref{fig6}. The strong competition between the dissociation and regeneration leads to a flat centrality dependence of $r_{AA}$. The calculations based on the transport approach agree well with the SPS and RHIC data at mid rapidity. In conclusion, the newly defined $p_T$ nuclear modification factor $r_{AA}$ for $J/\psi$ is sensitive to the nature of the fireballs formed in heavy ion collisions. At mid rapidity it is dramatically and qualitatively different at SPS, RHIC and LHC,
\begin{equation}
\label{raa2}
r_{AA}\left\{ \begin{array}{ll}
>1 &\textrm{SPS}\\
\sim 1 &\textrm{RHIC}\\
<1 &\textrm{LHC}
\end{array} \right .
\end{equation}

Since there are currently no LHC data of $\langle p_T^2\rangle$ at mid rapidity, we turn to the calculation at forward rapidity. Taking into account the fact that the hot nuclear matter effects at forward rapidity are weaker than that at mid rapidity, the LHC $r_{AA}$ shifts upwards at forward rapidity, and the RHIC one is similar to the SPS one at mid rapidity. Again the theory calculations are supported by the RHIC and LHC data. Since $\langle p_T^2\rangle$ is a normalized quantity, the bands at LHC due to the uncertainty in the charm quark cross section are not so wide as $R_{AA}$ in Fig.\ref{fig6}.

For all the calculations above in A+A collisions, we have used the assumption of thermalized charm quarks, inspired from the mass dependence of $R_{AA}$ for different mesons and the sizeable D meson flow. What is the case if charm quarks are not thermalized? The charm quark motion in medium can be described by a Langevin equation~\cite{langevin}
\begin{equation}
\label{langevin}
{d{\bf p}\over dt}=-\gamma(T){\bf p}+{\bf \eta}(T),
\end{equation}
where ${\bf \eta}$ is a Gaussian noise variable and $\gamma$ the drag coefficient determined by the elastic collision processes and gluon radiation of charm quarks in the medium~\cite{zhuc,shanshancao}, both are functions of the medium temperature. We will not study here the details of the solution of the equation, instead we consider the other limit of the charm quark distribution: the pQCD distribution without any interaction with the medium. Since there is no energy loss, the charm quarks keep their pQCD distribution in the medium, and the regenerated charmonia will carry high momentum. In this case, the calculated $r_{AA}$ is much higher than the data~\cite{longpaper}. Even at forward rapidity, it is larger than unit at LHC. Since the thermal distribution can well describe the $R_{AA}$ and $r_{AA}$, we believe that charm quarks are thermalized at LHC. The other strong support to the charm quark thermalization is the observed $J/\psi$ flow at LHC~\cite{lhcjpsiv2,longpaper}. Without charm quark thermalization, there is no way to
create a sizeable $J/\psi$ flow. Note that the difference induced by the two limits of the charm quark distribution is not so big at RHIC, because the regeneration is not yet the dominant source of charmonium production at RHIC. At SPS, the difference disappears.

\section{Conclusions}
\label{s5}
The charmonium suppression has long been considered as a signal of the quark-gluon plasma created in relativistic heavy ion collisions. However, there are two kinds of hot nuclear matter effects on the charmonium production, the dissociation and the regeneration. The two affect the charmonium yield in an opposite way, and the degree of the both increases with increasing colliding energy. Therefore, the cancelation between the two weakens the sensitivity of the charmonium yield to the properties of the hot medium. The case is, however, dramatically changed when we focus on the charmonium transverse momentum distribution. The two hot nuclear matter effects work in different transverse momentum regions. The dissociation suppresses mainly the initial hard component, and the regeneration enhances the soft component. When the colliding energy increases, the dominant production source changes from the hard process to the soft process. The speed of the change is controlled by the degree of the charm quark thermalization. If charm quarks are thermalized fast, the change becomes significant. Therefore, a dominant soft component can be taken as a clear signal of the regeneration at quark level, namely the signal of the quark-gluon plasma at RHIC and LHC energy.

To realize the above idea, we developed a detailed transport approach for charmonia in high energy nuclear collisions. The hot medium is described by ideal hydrodynamics, and the charmonium motion in the medium is governed by transport equations, including the dissociation and regeneration as lose and gain terms and cold nuclear matter effects as initial condition of the transport. By solving the two groups of coupled equations, we calculated the global and differential nuclear modification factors $R_{AA}(N_{part})$ and $R_{AA}(p_T)$, the transverse momentum distribution $dN_{J/\psi}/p_Tdp_T$, the second moment of the distribution $\langle p_T^2\rangle$ and especially the $p_T$ ratio $r_{AA}=\langle p_T^2\rangle_{AA}/\langle p_T^2\rangle_{pp}$ for $J/\psi$ in A+A collisions. From the comparison with the data, our main findings are: 1) The newly defined $p_T$ nuclear modification factor $r_{AA}$ is very sensitive to the hot mediums produced in heavy ion collisions. For instance, at mid rapidity it changes from larger than unit at SPS to around unit at RHIC and to less than unit at LHC. 2) The charm quarks are almost thermalized at LHC and RHIC.

\noindent {\bf Acknowledgement:} The work is supported by the NSFC
and the MOST under grant Nos. 11335005, 11221504, 2013CB922000, 2014CB845400, and the DOE under grant No.DE-AC03-76SF00098.


\end{document}